\documentclass{article}
\pdfoutput=1
\usepackage{authblk}
\providecommand{\keywords}[1]{\textbf{\textit{Keywords---}} #1}

\usepackage{graphicx}
\setkeys{Gin}{draft=false}% force the display of graphics, even for draft mode

\usepackage{subcaption}

\usepackage{xcolor}
\definecolor{dark-gray}{gray}{0.2}

\usepackage{times}

\usepackage{xspace}
\xspaceaddexceptions{]\}}

\usepackage{environ}

\usepackage{pgf}
\usepackage{pgfplots}
\usepackage{tikz}
\usetikzlibrary{positioning}
\usetikzlibrary{arrows}
\usepackage[hyphens]{url}
\usepackage[colorlinks,urlcolor=blue,linkcolor=dark-gray,citecolor=gray]{hyperref}

\usepackage{epsfig}

\usepackage{amsmath, amsthm, amssymb, mathtools}
\allowdisplaybreaks

\usepackage{fp}

\usepackage{algorithm}
\usepackage{algorithmic}

\usepackage{listings}

\usepackage{multicol}

\usepackage{multirow}

\usepackage{ifdraft}

\let\oldFootnote\footnote
\newcommand\nextToken\relax
\renewcommand\footnote[1]{%
    \oldFootnote{#1}\futurelet\nextToken\isFootnote}
\newcommand\isFootnote{%
    \ifx\footnote\nextToken\textsuperscript{,}\fi}

\newcommand{\email}[1]{\href{mailto:#1}{\nolinkurl{#1}}}

\makeatletter
\DeclareRobustCommand\join[2]{%
  \def\first{}%
  \def\list{#2}%
  \@for\reserved:=\list\do{%
    \ifx\first\empty%
      \def\first{done}%
    \else%
      #1%
    \fi%
    \reserved%
  }
}
\makeatother

\newcommand\copyrighttext{%
\footnotesize
To the extent possible under law, Matthieu Vergne has waived all copyright and related or neighboring rights to this technical report. For a more detailed description of this waiving, visit:\\
\url{https://creativecommons.org/publicdomain/zero/1.0/}
}
\newcommand\copyrightnotice{%
\begin{tikzpicture}[remember picture,overlay]
\node[anchor=south,yshift=30pt] at (current page.south) {\fbox{\parbox{\dimexpr\textwidth-\fboxsep-\fboxrule\relax}{\copyrighttext}}};
\end{tikzpicture}%
}

\begin{document}

\title{Gold Standard for Expert Ranking:\\A Survey on the XWiki Dataset}

\author[1,2]{Matthieu Vergne}
% \author[1]{Angelo Susi}

\affil[1]{
  Center for Information and Communication Technology, FBK-ICT\newline
  Via Sommarive, 18 I-38123 Povo, Trento, Italy\newline
  \email{vergne@fbk.eu}%
}
\affil[2]{
  Doctoral School in Information and Communication Technology\newline
  Via Sommarive, 5 I-38123 Povo, Trento, Italy\newline
  \email{matthieu.vergne@unitn.it}%
}

\date{2016-03-11}

\maketitle

%show to the reader the interest of the paper in 10 lines, highlighting each part of the title and the novel points
\begin{abstract}
% Context
We are designing an automated technique to find and recommend experts for helping in Requirements Engineering tasks, which can be done by ranking the available people by level of expertise.
% Problem
For evaluating the correctness of the rankings produced by the automated technique, we want to compare them to a gold standard.
% Proposal
In this work, we ask external people to look at a set of discussions and to rank their participants, before to evaluate the reliability of these rankings to serve as a gold standard.
% Contribution
We describe the setting and running of this survey, the method used to build the gold standard from the rankings of the subjects, and the analysis of the results to obtain and validate this gold standard.
% Results
Through the analysis of the results, we conclude that we obtained a reasonable gold standard although we lack evidences to support its total correctness.
We also made the interesting observation that the most reliable subjects build the least ordered rankings (i.e. has few ranks with several people per rank), which goes against the usual assumptions of Information Retrieval measures.
\end{abstract}

% put in abstract
\copyrightnotice
\section{Introduction}
\label{sec:introduction}

% Describe the context and important concepts
In Requirements Engineering (RE), we aim at managing the requirements of a project (i.e. the formalized needs of its stakeholders) in an effective and efficient way.
One task for this is to \emph{elicit} the requirements, which means to go to the stakeholders and identify their needs before to formalize them into specifications.
Another important task is to ensure that these requirements \emph{evolve} with the needs of the stakeholders, who can discover new constraints, face an evolving environment, or simply change their minds after receiving additional information.
In order to help in this requirements building and refinement, a high level of expertise is generally required in order to consider the multiple perspectives and their interdependencies.
Consequently, we focus on finding experts within a community of stakeholders, which is particularly relevant in Open Source projects having huge communities of diverse participants.
In particular, we are designing an automated technique to help finding the most expert participants, which can be done by ranking them by level of expertise~\cite{vergne_expert_2014}.

% Describe the problem and its consequences, highlighting each issue
To ensure that this automated technique works properly, we plan to compare it to a Gold Standard (GS), which should allow us to know, given a topic of expertise, how to rank the participants by decreasing expertise.
The issue here is that, for each community, the participants are different and the topics change, leading to the inability to provide a general GS applicable to any community.
This means that we need to have community-specific GSs, which build on the data available in this community to rank its participants.

% Describe the proposed solution, highlighting how it aims to tackle each issue
% (v) the structure of the paper
In this work, we target the XWiki community, which is described in more details in Section~\ref{sec:context} and is a large community composed of professional programmers, volunteer contributors, and simple users of the XWiki platform.
In order to build a GS for XWiki, we organised a survey involving people out of this community, and we asked them to look at our XWiki dataset to evaluate and rank their participants, as described in Section~\ref{sec:survey}.
We organised the survey in order to give enough flexibility to build partial rankings, and designed a method in Section~\ref{sec:gsBuilding} to infer the final GS based on the multiple rankings provided by the subjects of the survey.
The running of the survey is described in Section~\ref{sec:running} and its results are described and analysed in Section~\ref{sec:results}.
We enriched the survey with additional questions to assess the reliability of the subjects' rankings, which is of particular importance to validate the GS built from them.
All the data of this survey can be accessed freely online\footnote{Experiment access: \url{http://selab.fbk.eu/vergne/Experiment-2014-02-19/}}.

% (vi) (opt) suggestions for reading (notations, remarks, ...)

\section{XWiki Dataset}
\label{sec:context}

XWiki\footnote{XWiki platform: \url{http://dev.xwiki.org}} is an Open Source Software (OSS) which takes the form of a platform for managing wikis.
It has a community of contributors, including a company managing the development of the OSS and selling support and training on it.
This community interact through different media, in particular a mailing list for support and discussions about the software.

We have used the archives of the XWiki mailing list, which are freely available online\footnote{XWiki archives: \url{http://lists.xwiki.org/pipermail/users/}}, to retrieve the e-mails exchanged and re-build the discussion threads.
We have restricted to e-mails of the year 2012, and we removed the discussions started before 2012 to ensure having consistent threads.
Consequently, we retrieved 2728 e-mails organized in 713 threads, having each between 1 and 37 e-mails.
All of them have been organized and formatted in order to present them to human subjects\footnote{Survey threads: \url{http://selab.fbk.eu/vergne/Experiment-2014-02-19/dataset/}}.

To build our GS, we needed to identify the topics on which people should be ranked, what we did by searching for topics having a reasonable amount of information.
By reasonable, we mean:
\begin{itemize}
 \item[\emph{E-mail min}] having enough e-mails about the topic to ensure that the subjects have a high chance to obtain relevant information to evaluate the expertise of each participant,
 \item[\emph{E-mail max}] having small enough data to ensure that an average human can deal with it,
 \item[\emph{Thread average}] avoiding short discussions (i.e. 1-2 messages) which tend to remain superficial on the topic.
\end{itemize}
The exact limits cannot be decided a priori because (i) it should be balanced with the number of topics our subjects will have to work on, otherwise they could be overwhelmed by the amount of information to consider, and (ii) we did not know in advance how many topics could satisfy these requirements.
To satisfy them, we extracted the terms used in the e-mails to identify the available topics, and for each of them we counted how many threads are about them and how many e-mails they represent.
We did so automatically to have an approximative idea and finalized the selection manually, which lead us to select two topics (the lists of numbers provide the thread IDs in the dataset):
\begin{itemize}
  \item[\emph{Debian}] 34 e-mails in 6 threads: 71, 251, 546, 560, 562, 667
  \item[\emph{Hibernate}] 37 e-mails in 8 threads: 147, 153, 154, 172, 185, 444, 576, 687
\end{itemize}
Because we wanted all our subjects to deal with both topics in an hour, we evaluated that 30 to 40 e-mails per topic was a reasonable amount, and an average of 5 e-mails per thread was enough to have informative discussions.

\section{Survey Procedure and Material}
\label{sec:survey}

Following the terminology of \cite{wohlin_experimentation_2012}, the procedure described here is something between a survey and a quasi-experiment.
This is a survey in the sense that we aim at obtaining opinions from a population of subjects rather than checking some pre-defined hypothesis, but an experiment aspect is provided by the control variables we use to help analysing and validating the results (the GSs built).
The classification as a quasi-experiment rather than an experiment is due to the selection of the topics, which is not random.
Additionally, the fact that we do not use random subjects but volunteers from a specific community (mainly PhD students in RE) means that we make a \emph{convenience sampling}, which significantly reduces the randomness too.

The survey was organized in several phases:
\begin{enumerate}
 \item Present the survey to the subjects
 \item fill the pre-questionnaire
 \item Execute the main task on one of the two topics and fill a questionnaire
 \item Execute the main task on the other topic and fill another questionnaire
 \item fill the post-questionnaire
\end{enumerate}
The presentation provides a common perspective to all the subjects to work on their tasks and describe the survey process.
The pre-questionnaire focuses on the subject's profile and the post-questionnaire on the feedback about the tasks executed and the survey in general.
More details are given in the following subsections.

\subsection{Presentation: Common Subject Perspective}

In order to minimize the interpretation misalignments of the subjects, we gave them a common perspective by introducing a synthetic context\footnote{Survey presentation:\\\url{http://selab.fbk.eu/vergne/Experiment-2014-02-19/presentation.pdf}}.
This context was designed based on the expected profile of the subjects, mainly PhD students in RE.
Consequently, the context presented was to take the role of a requirement analyst in a small company in Information Technologies, aiming for refining a set of existing requirements.
For their imaginary job, they were asked to find people to help them obtain the relevant information about some topics related to the requirements to refine.
This is with this goal in mind that we asked them to rank XWiki participants by level of expertise, based on the intervention of these participants.

\subsection{Pre-Questionnaire: Subject Profiles}

The pre-questionnaire\footnote{Pre-questionnaire:\\\url{http://selab.fbk.eu/vergne/Experiment-2014-02-19/pre-questionnaire.pdf}} focuses on the profile of the human subject.
In particular, we asked their current position (e.g. undergraduate, PhD, professional) and how familiar they are with OSS in general and XWiki in particular.
Additionally, because we ask them to rank people by expertise, we also asked the subjects how familiar they are with the expert finding task.
We also asked how familiar they are with requirement analysis because it is the perspective they were asked to take (i.e. searching for experts to help them refining requirements).
We did not informed the subjects about the topics (i.e. Debian and Hibernate) before to give them the corresponding questionnaire, which is described below.

\subsection{Main Questionnaire: Expert Rankings}

The main task aims at searching for experts on a given topic, Debian or Hibernate, by looking at the e-mails of the XWiki participants.
Subjects are given one of the two topics\footnote{Debian questionnaire:\\\scriptsize\url{http://selab.fbk.eu/vergne/Experiment-2014-02-19/questionnaire-debian.pdf}}\footnote{Hibernate questionnaire:\\\scriptsize\url{http://selab.fbk.eu/vergne/Experiment-2014-02-19/questionnaire-hibernate.pdf}} and are asked to search for relevant discussion threads on this topic and rank their participants by decreasing expertise.
Consequently, we asked the subjects to list the discussions they looked at and to rank their participants.
We also asked them about factors which could hurt the reliability of the subject's ranking: the expertise of the subject himself on the topic, the confidence he has in his ranking, and how difficult it was to build it.
Each subject has executed this task on both topics, half starting with Debian, the other half with Hibernate, so we can have enough data for each topic.
Swapping the starting topic between subjects can help to identify a learning effect, for instance by seeing if participants highly ranked on the first topic tend to be ranked higher in the second.

\begin{figure}
  \centering
  \fbox{\includegraphics[trim={20px 410px 30px 110px},clip,width=0.98\textwidth]{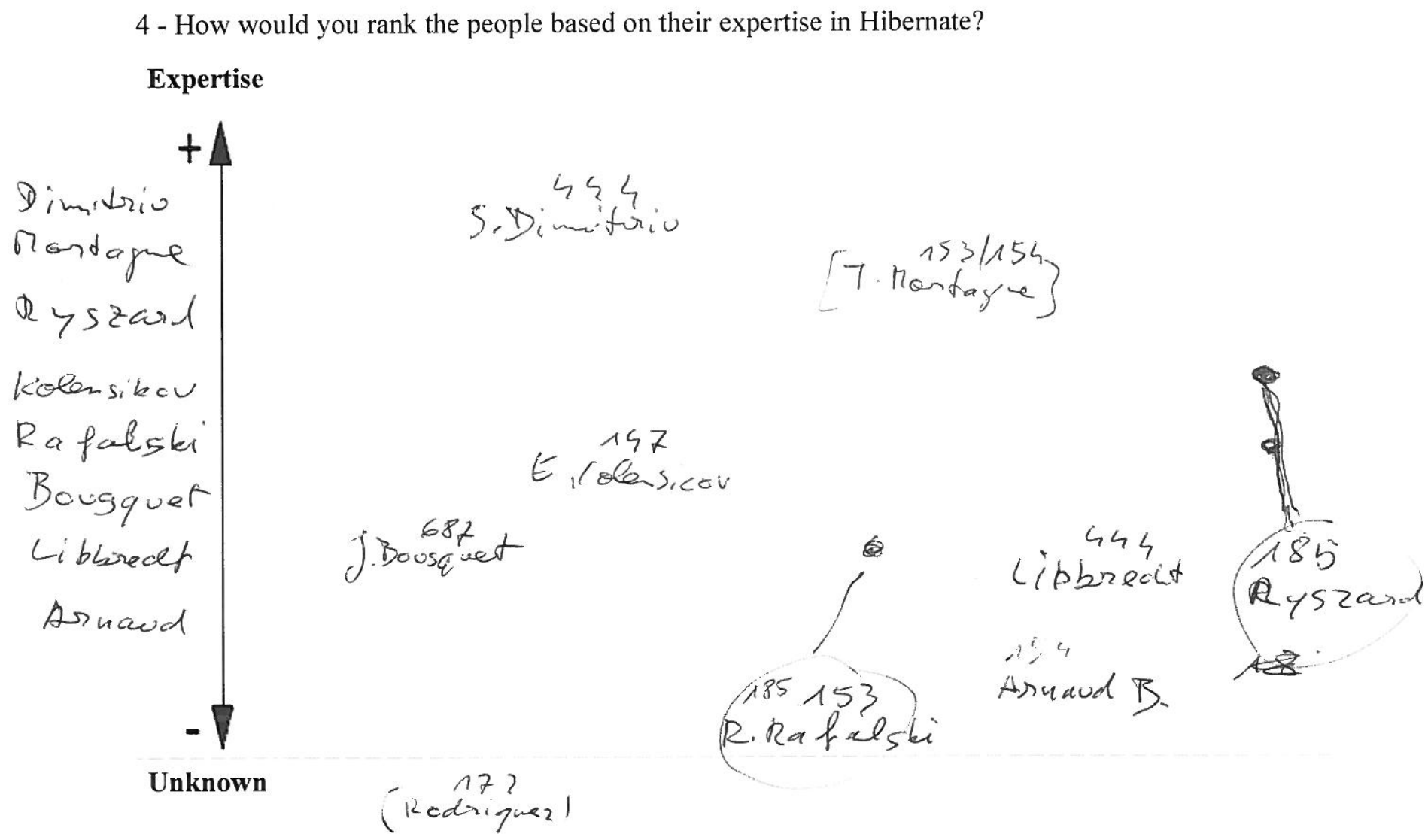}}
  \caption{Example of use of the ranking space: the right side of the scale is used to place and revise the position of each participant, while the left side summarizes the final ranking.}
  \label{fig:scan}
\end{figure}

In order to produce the rankings, a lot of flexibility was provided: a large blank area was available, with an arrow to show that the most experts should be placed at the top and the least experts at the bottom of this area.
This way, the subjects can place several people at the same level, allowing us to know when the subject has not enough information to tell which one is better, or to draw more complex structures if needed, as shown in Figure~\ref{fig:scan}.
During the presentation of the survey, the subjects have been explicitly requested to exploit the area in such a way if required.

\subsection{Post-Questionnaire: Feedback}

To ensure that the survey runs properly, it is important to know if something went wrong or if the subjects had any difficulty to execute the requested tasks.
Many issues can be managed on the fly by the survey manager, like answering questions about the survey or helping to access the online resources, but some issues might remain unnoticed and need to be requested explicitly to the subjects.
The post-questionnaire\footnote{Post-questionnaire:\\\url{http://selab.fbk.eu/vergne/Experiment-2014-02-19/post-questionnaire.pdf}} allows to trace such issues, making us able to consider them when evaluating the results of the survey.
We asked in particular about the perceived ability of the subject to achieve the requested tasks in the available time, the clarity of the requests, and the ability to use the provided resources properly.
A free comment area was also available for any feedback that the subject would like to share which was not part of the questionnaires.

We also used the post-questionnaire as an opportunity to obtain additional feedback to know how the subjects built their rankings.
In particular, we asked which messages were helpful or not, and to describe the methods used to rank the participants.
While the subjects rankings allow us to build a GS for evaluating our automated approach, the answers of these additional questions can be of interest for fixing or improving it.

\section{Gold Standard Building}
\label{sec:gsBuilding}

\subsection{Retrieval of Ordered Pairs}

\newcommand\order{ordered pair\xspace}
\newcommand\ranking{ranking\xspace}

\newcommand\superiorOrder{\emph{Superior}\xspace}
\newcommand\superiorOrderShort{\ensuremath{>}\xspace}
\newcommand\inferiorOrder{\emph{Inferior}\xspace}
\newcommand\inferiorOrderShort{\ensuremath{<}\xspace}

\newcommand\supO[1]{\ensuremath{\join{\superiorOrderShort}{#1}}}
\newcommand\infO[1]{\ensuremath{\join{\inferiorOrderShort}{#1}}}
\newcommand\undO[1]{\ensuremath{[\join{, }{#1}]}}

\newcommand{\rankingMath}{\ensuremath{r}\xspace}
\newcommand{\rankingSet}{\ensuremath{R}\xspace}

\begin{figure}
  \begin{center}
    \begin{tikzpicture}[
	element/.style={},
	arrow/.style={->},
	scale=2,
      ]
      \coordinate (undecided) at (0,0);
      \coordinate (superior) at (1,0);
      \coordinate (inferior) at (0,1);
      
      \node at (undecided) {$+$};
      \node at (superior) {$+$};
      \node at (inferior) {$+$};
      
      \node[below left] at (undecided) {\begin{tabular}{r} $(0,0)$ \\ No order \end{tabular}};
      \node[below right] at (superior) {\begin{tabular}{l} $(1,0)$ \\ $a>b$ \end{tabular}};
      \node[above left] at (inferior) {\begin{tabular}{r}$a<b$ \\ $(0,1)$ \end{tabular}};
      
      \draw[arrow] (0,0) -- (1.2,0);
      \draw[arrow] (0,0) -- (0,1.2);
      \draw (undecided) -- (inferior) -- (superior) -- (undecided);
      \draw[dashed] (0,0) -- (0.5,0.5);
      
      \node at (0.75,0.1) {$(a)$};
      \node at (0.1,0.75) {$(b)$};
      
      \def\x{0.5}
      \def\y{0.2}
      \coordinate (example) at (\x,\y);
      \fill (example) circle (1pt);
      \draw (0,0) -- (example);
      \draw[dotted] (0,\y) node[left] {$y$} -- (example);
      \draw[dotted] (\x,0) node[below] {$x$} -- (example);
    \end{tikzpicture}
  \end{center}
  \caption{Distribution of the three cases of {\order}s in a 2D space. Falling in the area (a) leads to use a $a>b$ for the centroid, (b) leads to $a<b$, and the dashed line leads to no order. An example of vector $(x,y)$ falls in the area (a), thus being interpreted as $a>b$.}
  \label{fig:orderSpaceSquare}
\end{figure}
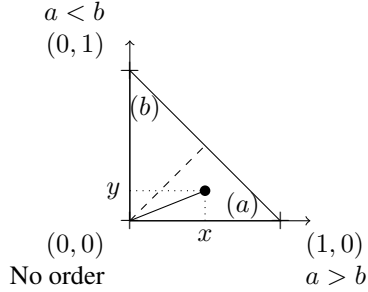

In this survey, several subjects provide rankings for each topic, leading to a set of rankings $\rankingSet = \{r_1, ..., r_n\}$ for each topic which need to be translated into a single ranking $\hat{\rankingMath}$ acting as a GS.
This single ranking is built by considering, for each pair of participants $(a,b)$, the most probable order ($a>b$ or $a<b$) depending on the different {\ranking}s in {\rankingSet}.
In such a way, we build a \emph{centroid} for {\rankingSet}, meaning a ranking which is ``in the center'' of {\rankingSet}.
To compute the {\order} of a given pair $(a, b)$, a 2D vector representation is used with Euclidian coordinates $(x,y)$, such that $x, y \in [0;1]$.
In particular, as illustrated in Figure~\ref{fig:orderSpaceSquare}, we associate a specific vector to each case of {\order}:
\begin{itemize}
 \item $a > b \Rightarrow (1, 0)$
 \item $a < b \Rightarrow (0, 1)$
 \item $\text{no order} \Rightarrow (0, 0)$
\end{itemize}

The last case occurs when $a$ or $b$ (or both) are not present in the ranking, so no order can be considered.
To identify the centroid {\order} for $(a, b)$, we compute a weighted average of these vectors, with the weights corresponding to the number of times they appear in {\rankingSet}.
More formally, for a set $R$ of $n$ {\ranking}s, $n_{s}$ {\ranking}s return $a>b$, $n_{i}$ return $a<b$, and $n_{u}$ return no order, with $n_{s} + n_{i} + n_{u} = n$.
We compute the average vector $(x,y)$ such that $x = \frac{n_{s}}{n}$ and $y = \frac{n_{i}}{n}$, which makes it falls between the three cases, as illustrated in Figure~\ref{fig:orderSpaceSquare}, and use the closest order ($>$ or $<$) to obtain the centroid {\order}.
In the case of perfect balance ($x = y$), no order is considered for the centroid.

The interpretation behind this model is that, if a majority of {\ranking}s agree on having a given {\order}, this {\order} will be the one used for the centroid.
We might argue that the ``no order'' case is not well represented and should share the 2D area with the two other cases but, as we show in the next section, this leads to a loss of information translated into arbitrary ordered pairs.
Consequently, we prefer to reduce the ``no order'' case to the minimum and favour the two other cases to preserve as much original ordered pairs as possible.

It is worth noting that, because we consider the ordered pairs independently, the transitivity of the rankings is not necessarily preserved in the centroid pairs.
Indeed, by having the {\ranking}s $r_1 = \supO{a,b,c}$, $r_2 = \supO{c,a,b}$, and $r_3 = \supO{b,c,a}$, we obtain the centroid pairs $a>b$, $b>c$, and $c>a$, so a loop.
To obtain a proper ranking, we need to restore the transitivity property of these pairs, process that we describe in the next section.

\subsection{Restoring the Transitivity Property}

By retrieving the ordered pairs separately, we do not consider their dependencies, leading to a set of ordered pairs which do not correspond to a proper ranking (i.e. the transitivity property is not satisfied).
In order to fix it, we use Algorithm~\ref{algo:ordering2Ranking} which can be summarized in 4 steps:
\begin{enumerate}
 \item[(1-9)] retrieve all the ordered pairs $a>b$,
 \item[(11-15)] add the transitive ordered pairs ($\supO{a,b} \wedge \supO{b,c} \Rightarrow \supO{a,c}$),
 \item[(17-21)] remove the loops ($\supO{a,b} \wedge \supO{b,a} \Rightarrow \text{no order for $a$ and $b$}$),
 \item[(23-32)] build a {\ranking} by looking iteratively for dominant participants.
\end{enumerate}

\begin{algorithm}
\begin{algorithmic}[1]
  \REQUIRE $pairs$: {\order}s for the centroid
  \ENSURE $\hat{r}$: {\ranking} built
  \STATE $SUP \leftarrow \emptyset$
  \STATE $E \leftarrow elementsOf(pairs)$
  \FORALL{$(a, b) \in E \times E$}
    \IF{$a>b \in pairs$}
      \STATE $SUP \leftarrow SUP \cup \{(a,b)\}$
    \ELSIF{$a<b \in pairs$}
      \STATE $SUP \leftarrow SUP \cup \{(b,a)\}$
    \ENDIF
  \ENDFOR
  \STATE 
  \FORALL{$(a, b, c) \in E \times E \times E$}
    \IF{$\{(a, b), (b, c)\} \subset SUP$}
      \STATE $SUP \leftarrow SUP \cup \{(a, c)\}$
    \ENDIF
  \ENDFOR
  \STATE 
  \FORALL{$(a, b) \in E \times E$}
    \IF{$\{(a, b), (b, a)\} \subset SUP$}
      \STATE $SUP \leftarrow SUP \backslash \{(a, b), (b, a)\}$
    \ENDIF
  \ENDFOR
  \STATE 
  \STATE $\hat{r} \leftarrow \emptyset$
  \STATE $rank \leftarrow 0$
  \WHILE{$||SUP|| > 0$}
    \STATE $top \leftarrow \{e \in E | \exists x \in E, (e,x) \in SUP \wedge \nexists y \in E, (y,e) \in SUP\}$
    \FORALL{$e \in top$}
      \STATE $\hat{r}(e) \leftarrow rank$
    \ENDFOR
    \STATE $SUP \leftarrow SUP \backslash \{(e,x) \in SUP | e \in top\}$
    \STATE $rank \leftarrow rank + 1$
  \ENDWHILE
\end{algorithmic}
\caption{Procedure used to build a ranking from a set of ordered pairs.}
\label{algo:ordering2Ranking}
\end{algorithm}

The first phase retrieves the explicit data, the second phase infers the implicit one, the third phase resolves the over-constrained pairs, and the last phase resolves the under-constrained ones (add arbitrary {\order}s to produce a proper ranking).
In particular, during this last phase, if the information inferred so far shows that \supO{a,b,c} and \supO{d,e,f,g}, without having any information between the elements of the two subsets, then the final ranking arbitrarily merges them into ${\rankingMath} = \supO{\undO{a,d},\undO{b,e},\undO{c,f},g}$.
Even if some relations occur, like for \supO{a,x,b,c} and \supO{d,e,x,f}, the final ranking arbitrarily merges them into ${\rankingMath} = \supO{\undO{a,d},e,x,\undO{b,f},c}$, while it could have been ${\rankingMath} = \supO{d,\undO{a,e},x,b,\undO{f,c}}$ as well as many others.
These arbitrary choices having an impact on how the stakeholders are ranked (so who we consider as more expert), it is important to obtain sufficient information to be able to build a proper ranking (at least for the top stakeholders).
This is why we reduce, in the previous section, the ``no order'' case to a single line rather than a 2D area.

\section{Survey Execution}
\label{sec:running}

For running the survey, we have invited by e-mail people from a RE research group to participate as volunteers (no incentive was provided).
10 people have accepted to participate in the survey.
The plan of the survey has been followed rigorously, starting from the presentation of the survey to the subjects (no detail have been given in the initial invitation to avoid any preparation).
The common perspective has been described and the dataset and questionnaires have been presented, showing how the task can be executed (a different topic have been used for the presentation to not bias the subjects).
Once the pre-questionnaires have been filled, all the subjects have received their first main questionnaire at the same time.
The task has last 20 minutes and we notified them 5 minutes before the end so they could properly finish their task.
Once done, the questionnaires of the first task have been exchanged with new questionnaires for the second task (i.e. for the other topic) and the same process of 20 minutes occurred.
Once finished, the questionnaires have been exchanged with the post-questionnaires, and the subjects were free to leave once their post-questionnaire was filled.

No significant issue was noticed during the execution of the survey, but the feedback of the post-questionnaires highlights some issues which may have significant impacts on the reliability of the rankings produced.
The most important issue seems to be the lack of time, which made it hard to read twice the e-mails, refine the rankings, or even consider all the relevant discussions.
A subject also checked first on Wikipedia (the survey was run with the online dataset, so they had access to Internet) to know better about the two topics before to work on the task, which means that even less time was available for this subject.
Another issue was the doubt the subjects could have in using the right criteria to rank properly the participants, or the lack of specificity of the topics, which decreases the confidence that subjects have in their rankings.
More superficial issues were mentioned, like using the names of the participants rather than short IDs makes it harder to rank them, or the fact that a subject was bored by the presentation.
In short, the main issues seem to be (i) some rankings are based on less information than others, and (ii) subjects may lack in confidence in their rankings.
Putting aside these free comments, the whole analysis of the survey provide additional insights which are given in the next section.

\section{Survey Results}
\label{sec:results}

The complete analysis of the questionnaires is done in this section.
We first analyse the subjects who participated through their answers to the pre-questionnaire in Section~\ref{sec:results-pre}.
Then, we introduce a common ground to the two main tasks in Section~\ref{sec:results-common} before to go in a deep analysis of each task in the sections \ref{sec:results-Debian} and \ref{sec:results-hibernate}, where we identify the GSs for each topic and evaluate their reliability.
We conclude this section by analysing the feedback given in the post-questionnaires in Section~\ref{sec:results-post} and list the main threats to validity.

\subsection{Subjects' Profiles}
\label{sec:results-pre}

10 people have participated as subjects for the survey: 1 undergraduate and professional, 8 PhD students, and 1 researcher.
As expected, all of them are familiar with requirement analysis methods (4 have worked with some, 6 are used to apply them), so none of them should have difficulties to act based on the common context given during the presentation.
However, none of them is familiar with the expert finding task (6 never did it, 4 did it without applying any method), which means that not only it could be difficult for them to rank the people, but they could also use wrong ranking criteria.
None of them are familiar with the XWiki dataset used (7 did not know about XWiki, 3 only heard about it) which means that the rankings produced should have no bias due to initial knowledge of the subjects on this project.
Finally, although they don't know about XWiki, some of them are familiar with Open Source Software (5 have produced OSS code or participated in OSS communities), still half of them are not familiar and so could have additional difficulties to understand the discussions.

\subsection{Main Tasks: Common Grounds}
\label{sec:results-common}

By summing both topics (Debian and Hibernate), the rankings have been produced based on 14 discussion threads containing 71 e-mails written by 18 participants.
For facilitating the redaction of this report, we assign each participant an ID:
\begin{multicols}{2}
\begin{enumerate}
 \item Adrian Fita
 \item Arnaud Bourree
 \item Nicolas Cheneau-Grehalle
 \item Denis Gervalle
 \item Eugene Colesnicov
 \item Guillaume Fenollar
 \item Jeremie Bousquet
 \item Marius Dumitru Florea
 \item Markus Kalkbrenner
 \item Matt Hammond
 \item Paul Libbrecht
 \item Philippe Marzouk
 \item Richard Hierlmeier
 \item Richard Rafalski
 \item Sergiu Dumitriu
 \item Thomas Mortagne
 \item Ricardo Rodriguez
 \item Ryszard Lach
\end{enumerate}
\end{multicols}

\subsection{Debian Rankings}
\label{sec:results-Debian}

For the topic Debian, 6 discussion threads were concerned (71, 251, 546, 560, 562, 667), with a total of 34 e-mails written by 13 participants (1, 2, 4, 5, 6, 8, 9, 10, 12, 13, 14, 15, 16).
Among the 10 subjects, 4 have considered all the threads, 2 have missed 1 thread (4-5 e-mails), and 4 have missed 2 threads (9-10 e-mails).
Over the 6 who missed threads, 4 were the first task, which might support a \emph{warm up effect}: the people who deal with the topic in second might have ``warmed up'' while treating the other topic first, leading to a better performance with the second topic.

% Discussions considered:
% \begin{itemize}
%  \item[] 71, 251, 546, 560, 562, 667
%  \item[] 71, 251, 546, 560, 562, 667
%  \item[] 71, 251, 546, 560, 562, 667
%  \item[] 71, 251, 546, 560, 562, 667
%  \item[1] 71, 251, 546, 560, 562, --- (4 missed)
%  \item[11] 71, 251, 546, 560, ---, 667 (5 missed)
%  \item[4] 71, 251, 546, 560, ---, --- (9 missed)
%  \item[9] 71, 251, 546, 560, ---, --- (9 missed)
%  \item[10] 71, 251, 546, 560, ---, --- (9 missed)
%  \item[7] 71, 251, 546, ---, ---, 667 (10 missed)
% \end{itemize}

The rankings produced by subjects working on Debian for their first task are the following:
\begin{itemize}
  \item[Subject 4] $[8]>[6, 13]>[16]>[14]>[10]$
  \item[Subject 5] $[8]>[15]>[6]>[4]>[16]>[2]>[9]>[13, 14]>[10]>[12]>[5]$
  \item[Subject 7] $[8, 13]>[6]>[16]>[1, 2]>[14]>[9]$
  \item[Subject 9] $[8]>[4, 6]>[16, 13, 15]>[1]>[14]>[10]>[12]$
  \item[Subject 11] $[1, 16, 4]>[6]>[9, 13]>[10, 15]>[14]$
\end{itemize}
and these ones for the second task:
\begin{itemize}
  \item[Subject 1] $[16]>[4, 8, 13]>[6, 15]>[2, 12]>[1]>[5, 9, 10, 14]$
  \item[Subject 3] $[1, 14]>[4, 8, 9, 10, 13]>[16, 2, 6]>[5, 12, 15]$
  \item[Subject 6] $[16]>[4, 6, 8, 13, 14, 15]>[1, 2, 9, 10, 12]>[5]$
  \item[Subject 8] $[16]>[4, 8]>[2, 12, 13, 15]>[6, 9, 14]>[1, 5, 10]$
  \item[Subject 10] $[16]>[6, 13, 14]>[4]>[1, 8, 10, 12, 15]$
\end{itemize}

An obvious difference appears between the two cases here.
With the first task, subjects mainly consider that the best expert for Debian is 8, and the fact that the only ranking not putting it first does not even consider it allows to say that it is a broad agreement.
At the opposite, when Debian is the second task, almost all the rankings agree that 16 should be considered as the top expert, although for the first task it was prone to be ranked in the middle.

We could consider several explanations for this significant change of the rank of 16.
The first explanation is that people having worked on Hibernate first (detailed in the Hibernate section) could have been influenced by the information learned about Hibernate.
Indeed, for both first and second task, Hibernate rankings put at unanimity 16 as top expert, leading to think that some strong evidences makes everyone agreeing on this perception.
Such a strong evidence might have influenced the subjects to put 16 as top expert also for Debian, especially if we interpret its middle location in the first task as poor evidence of high nor low expertise.
Another explanation is that, rather than a matter of influence, 16 could have provided additional information on his broad experience, including Debian, in discussions about Hibernate.
This explanation might be supported by the fact that the only ranking putting 16 on top for the first Debian task is made by the only professional subjects, so we might wonder whether the professional experience was helpful to analyse the discussions more efficiently, while others might have needed additional information to assess the expertise of 16.
The fact that some subjects mentioned not having enough time to revise their judgements could also be related to this case.
Unfortunately, the small data we have does not allow us to favour one explanation over the other.

For building the GS, we use the procedure described in Section~\ref{sec:gsBuilding} based on the rankings provided by the subjects.
In our case, we can build 3 GSs for Debian:
\begin{itemize}
 \item[First task] $[8]>[4, 6]>[13, 15, 16]>[1, 2]>[9]>[14]>[10]>[12]>[5]$
 \item[Second task] $[16]>[4, 13]>[8, 14]>[6]>[15]>[2]>[12]>[1]>[9]>[10]>[5]$
 \item[Both tasks] \small $[8]>[16]>[4]>[13]>[6]>[15]>[2]>[1]>[14]>[9]>[10]>[12]>[5]$
\end{itemize}
We can see, through the ranks of the participants 8 and 16, how the GS based on the rankings of the first task differs from the GS based on the rankings of the second task.
The last GS, based on both, merges these perspectives by having both participants on top.

These GSs are built from a set of rankings with the aim of reducing all these rankings to a single one, meaning that the GS should represent at best these rankings.
In order to assess this representativeness, we can measure the amount of agreement between the rankings of the set and the GS built from them.
More formally, we can decompose a ranking of the set and the corresponding GS into sets of ordered pairs and count how many pairs are in the same order ($a>b$ for both or $a<b$ for both), in the opposite order ($a>b$ for one and $a<b$ for the other), or in an unspecified state (at most one of them gives an order).

If we compare the rankings of the first task with their GS (Table~\ref{tab:agreementDebian}), we can see that the disagreement is always close to 0\%, showing that everyone is well represented.
We can see that the amount of unspecified is generally high, but this is due to the incompleteness of the rankings (e.g. the subject 4 ranks only 6 participants over 13) and their partial ordering (e.g. the subject 11 ranks 9 participants in only 5 ranks).
It is worth noting that no subject ranking is complete: with 13 participants in total, the subject rankings have between 6 and 12 participants, with an average of 9 participants per ranking.

For the second task, only the subject 10 is incomplete (10 participants), but the partial ordering is still relevant: although 13 participants are ranked, they are distributed in 4 to 6 ranks only.
The completeness of the rankings explains why the unspecified value is significantly lower than for the first task, while the partial ordering explains why it remains still far from 0\%.
Regarding the disagreement, we also have low values excepted for 1 ranking, which is the only ranking not having 16 at the top.

If we do not make the difference between the first and second task, and compare all the rankings to the global GS, we do not see significant changes.
The different values vary generally with a small amount (around 3 units), so the observations made by separating the tasks remain the same in the global case.

\begin{table}
\begin{center}
\begin{tabular}{|c|c|c|c|}
\hline
Subject & Agreement & Disagreement & Unspecified \\
\hline
\multicolumn{4}{|c|}{Comparison with first task GS} \\
\hline
4       & 13 (17\%) & 0 (0\%)      & 65 (83\%)   \\% $[8]>[6, 13]>[16]>[14]>[10]$
5       & 57 (73\%) & 4 (5\%)      & 17 (22\%)   \\% $[8]>[15]>[6]>[4]>[16]>[2]>[9]>[14, 13]>[10]>[12]>[5]$
7       & 23 (29\%) & 2 (3\%)      & 53 (68\%)   \\% $[8, 13]>[6]>[16]>[1, 2]>[14]>[9]$
9       & 41 (53\%) & 0 (0\%)      & 37 (47\%)   \\% $[8]>[4, 6]>[13, 15, 16]>[1]>[14]>[10]>[12]$
11      & 21 (27\%) & 6 (8\%)      & 51 (65\%)   \\% $[1, 4, 16]>[6]>[9, 13]>[15, 10]>[14]$
\hline
\multicolumn{4}{|c|}{Comparison with second task GS} \\
\hline
1       & 61 (78\%) & 5  (6\%)     & 12 (15\%)   \\% $[16]>[4, 8, 13]>[6, 15]>[2, 12]>[1]>[5, 14, 9, 10]$
3       & 35 (45\%) & 25 (32\%)    & 18 (23\%)   \\% $[1, 14]>[4, 8, 9, 13, 10]>[2, 6, 16]>[5, 12, 15]$
6       & 53 (68\%) & 0  (0\%)     & 25 (32\%)   \\% $[16]>[4, 6, 14, 8, 13, 15]>[1, 2, 9, 12, 10]>[5]$
8       & 55 (71\%) & 8  (10\%)    & 15 (19\%)   \\% $[16]>[4, 8]>[2, 12, 13, 15]>[6, 14, 9]>[1, 5, 10]$
10      & 27 (35\%) & 3  (4\%)     & 48 (62\%)   \\% $[16]>[6, 14, 13]>[4]>[1, 8, 12, 15, 10]$
\hline
\multicolumn{4}{|c|}{Comparison with both tasks GS} \\
\hline
1       & 62 (79\%) & 5  (6\%)     & 11 (14\%)   \\% $[16]>[4, 8, 13]>[6, 15]>[2, 12]>[1]>[5, 14, 9, 10]$
3       & 36 (46\%) & 25 (32\%)    & 17 (22\%)   \\% $[1, 14]>[4, 8, 9, 13, 10]>[2, 6, 16]>[5, 12, 15]$
4       & 12 (15\%) & 2  (3\%)     & 64 (82\%)   \\% $[8]>[6, 13]>[16]>[14]>[10]$
5       & 54 (69\%) & 11 (14\%)    & 13 (17\%)   \\% $[8]>[15]>[6]>[4]>[16]>[2]>[9]>[14, 13]>[10]>[12]>[5]$
6       & 50 (64\%) & 3  (4\%)     & 25 (32\%)   \\% $[16]>[4, 6, 14, 8, 13, 15]>[1, 2, 9, 12, 10]>[5]$
7       & 24 (31\%) & 2  (3\%)     & 52 (67\%)   \\% $[8, 13]>[6]>[16]>[1, 2]>[14]>[9]$
8       & 55 (71\%) & 10 (13\%)    & 13 (17\%)   \\% $[16]>[4, 8]>[2, 12, 13, 15]>[6, 14, 9]>[1, 5, 10]$
9       & 38 (49\%) & 3  (4\%)     & 37 (47\%)   \\% $[8]>[4, 6]>[13, 15, 16]>[1]>[14]>[10]>[12]$
10      & 22 (28\%) & 10 (13\%)    & 46 (59\%)   \\% $[16]>[6, 14, 13]>[4]>[1, 8, 12, 15, 10]$
11      & 24 (31\%) & 7  (9\%)     & 47 (60\%)   \\% $[1, 4, 16]>[6]>[9, 13]>[15, 10]>[14]$
\hline
\end{tabular}
\end{center}
\caption{Amount of agreements between the subjects' rankings and the GSs based on them for the topic Debian.}
\label{tab:agreementDebian}
\end{table}

Additionally, we can compare the GSs between each other (Table~\ref{tab:agreementDebianGS}).
The small differences between the task-specific comparison and the global comparison let imagine that the task-specific GSs are quite close to the global one, which is confirmed by their low disagreements (5\% and 13\%).
Naturally, the global GS being a trade-off between the two tasks, by summing their disagreements with the global GS we retrieve the disagreement between both (18\%).
The biggest difference between the GSs and their rankings is that, because each GS brings as much information as possible from its rankings, they tend to be complete (all participants are ranked) and totally ordered (as many ranks than participants), although it is not guaranteed.
This is why the amount of unspecified is close to 0\% when comparing GSs.

\begin{table}
\begin{center}
\begin{tabular}{|c|c|c|c|}
\hline
GSs              & Agreement & Disagreement & Unspecified \\
\hline
Both vs. first   & 69 (88\%) & 4  (5\%)     & 5 (6\%)     \\
Both vs. second  & 66 (85\%) & 10 (13\%)    & 2 (3\%)     \\
First vs. second & 57 (73\%) & 14 (18\%)    & 7 (9\%)     \\
\hline
\end{tabular}
\end{center}
\caption{Amount of agreements between the Debian GSs.}
\label{tab:agreementDebianGS}
\end{table}

Finally, our aim being of producing a reliable GS for Debian, we need to evaluate the reliability of our 3 GSs.
For this, we can look at the perception of the subjects (Table~\ref{tab:perceptionDebian}), from who we asked to evaluate their own level of expertise on the topic, their confidence in the ranking they have produced, and how difficult it was to produce it.
Basically, someone having a high level of expertise, a high level of confidence, and a low level of difficulty should be particularly well represented by our final GS.

From the first task, the subjects 4 and 5 are the highest experts (4/5), and 4 in particular also has a high confidence (4/5) and a low difficulty (2/5).
By looking at the agreement between 4 and the GSs (Table~\ref{tab:agreementDebian}), we see that it is indeed perfectly represented by the first GS (0 disagreement) and has the lowest disagreement with the global one.
The main issue with 4 is that his ranking ranks only 6 participants over 13, so it does not provide a lot of information.

By considering the second task, only the subject 6 seems to stand out through his expertise, reinforced by a perfect confidence (5/5) and no difficulty (1/5).
Once again, Table~\ref{tab:agreementDebian} shows that this subject is perfectly represented with its task-specific GS (0 disagreement) and is among the best at the global level (only 3 disagreements).
However, if he does not suffer from the incompleteness issue of the previous expert, he suffers from the partial ordering: only 4 ranks to order 13 participants, which means that it also lacks a lot of information.

As additional evidences, we might consider other subjects having high confidence (7 for first task, 3 for second), but their lack of expertise (1 or 2) decreases their reliability, and we can see from Table~\ref{tab:agreementDebian} that 7 is indeed among the closest to the GS but 3 is the farthest, with two or three times more disagreement than the second farthest.
Consequently, we cannot rely much on these evidences to strengthen our first perception, so we can only say that all the Debian GSs seem to be globally correct with some reserves on the details.
In such a situation, we might say that the global GS, which builds on a trade-off, might be the most reliable GS.

\begin{table}
\begin{center}
\begin{tabular}{|c|c|c|c|c|c|}
\hline
\multirow{2}{*}
{Subject} & Participants & Ranks     & Expertise & Confidence & Difficulty \\
          & ranked       & used      & (1-5)     & (1-5)      & (1-5)      \\
\hline
\multicolumn{6}{|c|}{First task} \\
\hline
4         & 6            & 5  (83\%) & 4         & 4          & 2          \\
5         & 12           & 11 (92\%) & 4         & 3          & 4          \\
7         & 8            & 6  (75\%) & 1         & 4          & 3          \\
9         & 10           & 7  (70\%) & 2         & 3          & 4          \\
11        & 9            & 5  (56\%) & 2         & 3          & 3          \\
\hline
\multicolumn{6}{|c|}{Second task} \\
\hline
1         & 13           & 6  (46\%) & 2         & 3          & 3          \\
3         & 13           & 4  (31\%) & 2         & 4          & 2          \\
6         & 13           & 4  (31\%) & 4         & 5          & 1          \\
8         & 13           & 5  (38\%) & 2         & 3          & 4          \\
10        & 10           & 4  (40\%) & 2         & 3          & 3          \\
\hline
\end{tabular}
\end{center}
\caption{Ranking properties and subjects' perception for Debian.}
\label{tab:perceptionDebian}
\end{table}

\subsection{Hibernate Rankings}
\label{sec:results-hibernate}

For the topic Hibernate, 8 discussion threads were concerned (147, 153, 154, 172, 185, 444, 576, 687), with a total of 37 e-mails written by 10 participants (2, 3, 5, 7, 11, 14, 15, 16, 17, 18).
Among the 10 subjects, 7 have considered all the threads, 1 have missed 2 threads (3 e-mails), and 2 have missed 3 threads (4-6 e-mails).
Over the 3 who missed threads, 2 were the first task and 1 the second, which seems to support a warm up effect although it is rather small (at least, it does not contradict it).
Additionally, among the 3 who missed threads for Hibernate, 2 of them also missed threads for Debian, but the number of missed threads for Debian is doubled (6), so if we might have some subjects who tend to be slower than others, the warm up effect still seems to be a relevant explanation (one of the subjects even confirmed that he re-used his rankings from the first task for the second).
It is also interesting to notice that, although Hibernate has more discussions and more e-mails, the subjects have been generally more efficient in this task, even when it was the first task.
The fact that less participants were involved could be an explanation, because it reduces the ranking effort, but it could also be that the discussions were easier to understand.

% Discussions considered:
% \begin{itemize}
%  \item[] 147, 153, 154, 172, 185, 444, 576, 687
%  \item[] 147, 153, 154, 172, 185, 444, 576, 687
%  \item[] 147, 153, 154, 172, 185, 444, 576, 687
%  \item[] 147, 153, 154, 172, 185, 444, 576, 687
%  \item[] 147, 153, 154, 172, 185, 444, 576, 687
%  \item[] 147, 153, 154, 172, 185, 444, 576, 687
%  \item[] 147, 153, 154, 172, 185, 444, 576, 687
%  \item[10] 147, 153, 154, 172, 185, 444, ---, --- (3 missed)
%  \item[8] 147, 153, 154, 172, 185, ---, ---, --- (6 missed)
%  \item[11] 147, 153, 154, ---, 185, 444, ---, --- (4 missed)
% \end{itemize}

The rankings produced by subjects working on Hibernate for their first task are the following:
\begin{itemize}
  \item[Subject 1] $[15, 16]>[5, 7, 11, 18]>[2, 3, 17]>[14]$
  \item[Subject 3] $[14, 16]>[7, 18]>[15]>[2, 3, 5, 11, 17]$
  \item[Subject 6] $[16]>[14]>[15, 18]>[2, 5, 11]>[3, 7, 17]$
  \item[Subject 8] $[16]>[2, 14, 17, 18]>[5]$
  \item[Subject 10] $[16]>[2, 5, 11, 14, 17, 18]$
\end{itemize}
and these ones for the second task:
\begin{itemize}
  \item[Subject 4] $[16]>[14, 15, 18]>[5, 7]>[2]>[3, 11, 17]$
  \item[Subject 5] $[16]>[15]>[14]>[2]>[18]>[7]>[3, 11]>[5, 17]$
  \item[Subject 7] $[16]>[11, 14, 15]>[2, 3, 5, 7, 17, 18]$
  \item[Subject 9] $[16]>[15]>[14]>[11]>[18]>[17]>[2]>[3, 7]>[5]$
  \item[Subject 11] $[16]>[11]>[14]>[5, 18]>[2]$
\end{itemize}

At the opposite of Debian, no clear difference appear in the number of participants nor their order between the rankings of the first task and the rankings of the second task.
However, we observe a significant difference on the informativeness of these rankings: for the first task, the 10 participants are ranked on 2 to 5 ranks (3.6 in average), while the second task has between 3 and 9 ranks (6 in average).
Like for Debian, this difference can support a warm up effect, making the subjects more efficient on their second task.

For building the GS, we use the procedure described in Section~\ref{sec:gsBuilding} based on the rankings provided by the subjects.
Like for Debian, we can build 3 GSs for Hibernate:
\begin{itemize}
 \item[First task] $[16]>[14]>[15, 18]>[5, 7, 11]>[2]>[3, 17]$
 \item[Second task] $[16]>[15]>[14]>[11]>[18]>[2]>[7]>[3, 17]>[5]$
 \item[Both tasks] $[16]>[15]>[14]>[18]>[7, 11]>[2, 5]>[3, 17]$
\end{itemize}
Like for the subjects' rankings, there is few differences between the GSs, mainly the lack of information for the first task leading to have less ranks for its GS (6 ranks) compared to the second (9 ranks).
The global GS, naturally, makes a trade-off between the two.

If we compare the rankings of the first task to their GS (Table~\ref{tab:agreementHibernate}), we can see that the disagreement is, as expected, close to 0\%, excepted for the subject 1.
The high disagreement of this subject comes mainly from the participant 14, which is generally ranked high excepted for subject 1 who ranks her last.
We can also see that the subjects 8 and 10 have a high amount of unspecified, which is due in part to their incompleteness (6 or 7 participants over 10) but mainly to their reduced ordering (2 or 3 ranks, which is really poor in information).

If we look at the rankings of the second task, the subjects 7 and 11 also have a high amount of unspecified, 7 because of its partial ordering (only 3 ranks) and 11 mainly because of its incompleteness (6 participants over 10).
Still, the disagreement remains low, so the rankings are well represented by their GS.
Only the subject 4 reaches a high disagreement (18\%) because of the participants 5 and 11 which are swapped compared to the GS.

Although we observe a warm up effect, the main difference between the first and second task is the informativeness of their rankings.
Consequently, building a global GS makes more sense than for Debian, for which we saw significant differences in the orders.
By comparing all the rankings to this global GS, like for Debian, the values are not significantly impacted (around 2 units of difference in general).
This is coherent with the consistency we could observe between the rankings, consistency which is reflected in their GSs: if we compare the GSs between each other (Table~\ref{tab:agreementHibernateGS}), the disagreement is even closer to 0\% than for Debian.

\begin{table}
\begin{center}
\begin{tabular}{|c|c|c|c|}
\hline
Subject & Agreement & Disagreement & Unspecified \\
\hline
\multicolumn{4}{|c|}{Comparison with first task GS} \\
\hline
1       & 26 (58\%) & 8 (18\%)     & 11 (24\%)   \\% $[15, 16]>[5, 7, 11, 18]>[2, 3, 17]>[14]$
3       & 29 (64\%) & 1 (2\%)      & 15 (33\%)   \\% $[14, 16]>[7, 18]>[15]>[2, 5, 3, 11, 17]$
6       & 35 (78\%) & 1 (2\%)      & 9  (20\%)   \\% $[16]>[14]>[18, 15]>[2, 5, 11]>[7, 3, 17]$
8       & 7  (16\%) & 2 (4\%)      & 36 (80\%)   \\% $[16]>[2, 14, 17, 18]>[5]$
10      & 6  (13\%) & 0 (0\%)      & 39 (87\%)   \\% $[16]>[2, 5, 14, 11, 17, 18]$
\hline
\multicolumn{4}{|c|}{Comparison with second task GS} \\
\hline
4       & 30 (67\%) & 8 (18\%)     & 7  (16\%)   \\% $[16]>[14, 18, 15]>[5, 7]>[2]>[3, 11, 17]$
5       & 38 (84\%) & 4 (9\%)      & 3  (7\%)    \\% $[16]>[15]>[14]>[2]>[18]>[7]>[3, 11]>[5, 17]$
7       & 27 (60\%) & 0 (0\%)      & 18 (40\%)   \\% $[16]>[14, 11, 15]>[2, 5, 7, 3, 17, 18]$
9       & 41 (91\%) & 2 (4\%)      & 2  (4\%)    \\% $[16]>[15]>[14]>[11]>[18]>[17]>[2]>[7, 3]>[5]$
11      & 12 (27\%) & 2 (4\%)      & 31 (69\%)   \\% $[16]>[11]>[14]>[5, 18]>[2]$
\hline
\multicolumn{4}{|c|}{Comparison with both tasks GS} \\
\hline
1       & 27 (60\%) & 7 (16\%)     & 11 (24\%)   \\% $[15, 16]>[5, 7, 11, 18]>[2, 3, 17]>[14]$
3       & 29 (64\%) & 3 (7\%)      & 13 (29\%)   \\% $[14, 16]>[7, 18]>[15]>[2, 5, 3, 11, 17]$
4       & 34 (76\%) & 2 (4\%)      & 9  (20\%)   \\% $[16]>[14, 18, 15]>[5, 7]>[2]>[3, 11, 17]$
5       & 36 (80\%) & 4 (9\%)      & 5  (11\%)   \\% $[16]>[15]>[14]>[2]>[18]>[7]>[3, 11]>[5, 17]$
6       & 34 (76\%) & 3 (7\%)      & 8  (18\%)   \\% $[16]>[14]>[18, 15]>[2, 5, 11]>[7, 3, 17]$
7       & 25 (56\%) & 1 (2\%)      & 19 (42\%)   \\% $[16]>[14, 11, 15]>[2, 5, 7, 3, 17, 18]$
8       & 7  (16\%) & 1 (2\%)      & 37 (82\%)   \\% $[16]>[2, 14, 17, 18]>[5]$
9       & 35 (78\%) & 6 (13\%)     & 4  (9\%)    \\% $[16]>[15]>[14]>[11]>[18]>[17]>[2]>[7, 3]>[5]$
10      & 6  (13\%) & 0 (0\%)      & 39 (87\%)   \\% $[16]>[2, 5, 14, 11, 17, 18]$
11      & 11 (24\%) & 2 (4\%)      & 32 (71\%)   \\% $[16]>[11]>[14]>[5, 18]>[2]$
\hline
\end{tabular}
\end{center}
\caption{Amount of agreements between the subjects' rankings and the GSs based on them for the topic Hibernate.}
\label{tab:agreementHibernate}
\end{table}

\begin{table}
\begin{center}
\begin{tabular}{|c|c|c|c|}
\hline
GSs            & Agreement & Disagreement & Unspecified \\
\hline
Global vs. 1st & 38 (84\%) & 1 (2\%)      & 6 (13\%)    \\
Global vs. 2nd & 38 (84\%) & 4 (9\%)      & 3 (7\%)     \\
1st vs. 2nd    & 34 (76\%) & 6 (13\%)     & 5 (11\%)    \\
\hline
\end{tabular}
\end{center}
\caption{Amount of agreements between the Hibernate GSs.}
\label{tab:agreementHibernateGS}
\end{table}

Finally, our aim being of producing a reliable GS for Hibernate, we need to evaluate the reliability of our 3 GSs.
For this, like for Debian, we can look at the perception of the subjects (Table~\ref{tab:perceptionHibernate}) and rely on subjects having a high level of expertise, a high level of confidence, and a low level of difficulty.

For the first task, the most confident subject (subject 6) is also the least expert (1/5), so we might see (Table~\ref{tab:agreementHibernate}) that it is well represented by both its task-specific and global GS (1 disagreement each) but the lack of expertise make it unreliable to call it an evidence.
At the opposite, the most expert (subject 1) has a mitigated confidence (3/5) and difficulty (3/5), and seeing that it is the farthest from both its task-specific and global GS makes it even more questionable.

The second task is more interesting, because the highest expert (subject 7) is also the most confident (4/5) and among the ones having the least difficulty (2/5).
Like for Debian, we can assess that this subject's ranking is very well represented by both its task-specific GS (0 disagreement) and the global one (1 disagreement).
Still, like for Debian, this subject is from far the least informative with only 3 ranks to order 10 participants.

In short, we have less evidences than for Debian to confirm the reliability of our GSs, but the high similarity of all the rankings makes it less problematic.
Yet, we still see that the most expert and confident subjects have a tendency to have extremely partial rankings.

\begin{table}
\begin{center}
\begin{tabular}{|c|c|c|c|c|c|}
\hline
\multirow{2}{*}
{Subject} & Participants & Ranks     & Expertise & Confidence & Difficulty \\
          & ranked       & used      & (1-5)     & (1-5)      & (1-5)      \\
\hline
\multicolumn{6}{|c|}{First task} \\
\hline
1         & 10           & 4  (40\%) & 5         & 3          & 3          \\
3         & 10           & 4  (40\%) & 3         & 3          & 2          \\
6         & 10           & 5  (50\%) & 1         & 4          & 3          \\
8         & 6            & 3  (50\%) & 1         & 3          & 4          \\
10        & 7            & 2  (29\%) & 2         & 3          & 2          \\
\hline
\multicolumn{6}{|c|}{Second task} \\
\hline
4         & 10           & 5  (50\%) & 2         & 4          & 2          \\
5         & 10           & 8  (80\%) & 2         & 3          & 2          \\
7         & 10           & 3  (30\%) & 4         & 4          & 2          \\
9         & 10           & 9  (90\%) & 2         & 3          & 4          \\
11        & 6            & 5  (83\%) & 3         & 3          & 3          \\
\hline
\end{tabular}
\end{center}
\caption{Ranking properties and subjects' perception for Hibernate.}
\label{tab:perceptionHibernate}
\end{table}

\subsection{Feedback and Discussion}
\label{sec:results-post}

% Problems
From Table~\ref{tab:feedback}, we can see that the survey ran more or less smoothly.
In particular, the objectives, the main notions and the description of the tasks were clear, and the dataset was easy to use.
However, as mentioned earlier, the time was not sufficient for everyone to achieve the requested tasks and, although the dataset itself was easy to use, the relevant discussions were not easy to select nor understand.
These observations can provide an explanation to the difficulty values of 3 and more provided in the tables \ref{tab:perceptionDebian} and \ref{tab:perceptionHibernate}.

\begin{table}
\newcommand{\calc}[5]{\FPeval{\result}{round((1*#1+2*#2+3*#3+4*#4+5*#5)/10,1)}\result}
\begin{center}
\begin{tabular}{|l|c|c|c|c|c|c|}
\hline
\multirow{2}{*}
{Question} & \multicolumn{5}{|c|}{No \dotfill Yes} & \multirow{2}{*}{Avg}\\
           & 1 & 2 & 3 & 4 & 5 & \\
\hline
The time to perform the lab tasks was sufficient.
& 1 & 1 & 3 & 3 & 2 & \calc{1}{1}{3}{3}{2} \\
The objectives of the lab were clear.
& 0 & 0 & 0 & 5 & 5 & \calc{0}{0}{0}{5}{5} \\
The notion of “requirement analyst” was clear.
& 0 & 0 & 1 & 6 & 3 & \calc{0}{0}{1}{6}{3} \\
The notion of “expert finding” was clear.
& 0 & 1 & 0 & 3 & 6 & \calc{0}{1}{0}{3}{6} \\
The tasks were clear.
& 0 & 0 & 1 & 2 & 7 & \calc{0}{0}{1}{2}{7} \\
Using the dataset was easy.
& 0 & 0 & 3 & 3 & 4 & \calc{0}{0}{3}{3}{4} \\
The relevant discussions were easy to select.
& 0 & 2 & 4 & 1 & 3 & \calc{0}{2}{4}{1}{3} \\
The discussions I have read were easy to understand.
& 0 & 2 & 4 & 2 & 2 & \calc{0}{2}{4}{2}{2} \\
\hline
\end{tabular}
\end{center}
\caption{Feedback questions to evaluate how well the survey has run.}
\label{tab:feedback}
\end{table}

% Methods used
Additional questions of the post-questionnaire are also helpful to identify the properties which make a discussion relevant for building rankings.
In particular, the subjects highlighted the usefulness of discussions about problem resolutions and question-answers, but these types of discussions are the most common in our dataset, so other types of discussions might also be useful.
However, they also mentioned that clarification requests and messages with long logs are not helpful, probably because the content specific to the participant is minimal.
Naturally, long discussions were the most informative for the subjects, probably because they allow to work deeper on the problem/question, giving the opportunity to the participants to show better their expertise.
The explicit assessments of expertise, like self-assessment and recommendations of other people, were also considered by the subjects.

Once the relevant discussions are identified, the subjects built their rankings with their own strategies, several of them relying on the types of messages to evaluate who is a higher expert.
In particular, people providing answers were ranked higher than people asking, and people providing detailed messages were ranked higher than people writing short messages.
The number of messages was also a criteria, with more participation leading to a higher expertise.
More subjective criteria were also used to rank the participants, in particular the self-assessments and recommendations, but also the apparent confidence of the participant and the apparent broadness and depth of his knowledge.

A specific issue occurred from our side, in the analysis of the post-questionnaires.
Two questions were asked to the subjects, to know if they recognized some of the discussions and participants in the dataset.
These questions were asked with the intention to get more details from subjects already knowing about XWiki.
The issue was that, from the pre-questionnaire, 3 subjects only heard about XWiki, but in the post-questionnaire 2 subjects said that they recognized some of the discussions, which let think that they were more involved in the XWiki community than what they said in their pre-questionnaire.
Additionally, 5 said that they recognized some of the participants: although it might be that they know about some participants from contexts different to XWiki, it is still half of the subjects, and this possibility seems to us low enough to rise the flag.
Consequently, it might be that some misunderstandings occurred on these questions, for instance it might be that subjects mixed the recognition of participants in the \emph{discussions} with the recognition of participants in the \emph{survey}.
With such an interpretation, the results can be easily explained: most if not all the subjects know each other.
Due to this apparent contradictions, we let these answers out of our analysis.

% Threats to validity
Finally, based on all the results presented, we can discuss the validity of the survey in building reliable GSs.
Several observations might support a threat to the proper conduct of this survey (internal validity): the low confidence and high difficulty for some subjects to build their rankings, which can be due to the lack of time and the difficulty to select and understand the discussions, but also to the lack of expertise of the subjects in the topics.
The threats to the generalizability of our results (external validity) are obvious and numerous: few subjects, few topics, specific dataset, specific discussions, ranking based only on the participants of these discussions, etc. make our GSs highly specific to the context in which they have been built.
However, this is not an issue for us, because this survey was precisely intended to build a GS based on specific data, such that we can use this very same data in our automated technique and see how close it is from the GS.
In particular, by having subjects from the XWiki community, they would already have some knowledge to help them build their rankings: we preferred to have subjects out of the community to be closer to the situation of the automated technique, which does not have this initial knowledge.
We did not have an initial hypothesis to check, so we consider no threat to the construct validity, but the conclusion validity, that we link to the reliability of our GS, have some threats which deserve to be considered.
In particular, we saw that the most reliable subjects (high expertise, high confidence, low difficulty) were among the least informative (most incomplete or most partially ordered), thus giving only superficial support to confirm the reliability of our GSs.
Only the broad agreement of these GSs (how close they are to the rankings they are based on) supports their reliability, although it is only an evidence of agreement, not of correctness.

% Progress - describe the progress made in solving the stated problem and propose a plan to complete the research. The plan should include your strategy for evaluating your work and presenting credible evidence of your results to the research community.

\section{Conclusion}
\label{sec:conclusion}

% Context
We are designing an automated technique to find and recommend experts for helping in Requirements Engineering tasks, which can be done by ranking the available people by level of expertise.
% Problem
For evaluating the correctness of the rankings produced by the automated technique, we want to compare them to a gold standard.
% Proposal
In this work, we ask external people to look at a set of discussions and to rank their participants, before to evaluate the reliability of these rankings to serve as a gold standard.
% Contribution
We describe the setting and running of this survey, the method used to build the gold standard from the rankings of the subjects, and the analysis of the results to obtain and validate this gold standard.
% Results

% final results to keep in mind
Through this survey, we tried to build a gold standard to know how to rank people by decreasing expertise for a specific dataset (XWiki).
Through the analysis of the survey, we obtained a reasonable gold standard although we lack evidences to support fully its correctness.
We also made the interesting observation that the most reliable subjects build the least ordered rankings (i.e. has few ranks with several people per rank), which goes against the usual expectations for Information Retrieval measures.
This observation appears to us as an important one, because expert finding systems are mainly inspired from Information Retrieval systems~\cite{balog_expertise_2012}, where the ranking validation procedures are designed for complete and totally ordered gold standards~\cite{manning_introduction_2008}.

% (optionally) the future works.
Additionally, it might be interesting to investigate further the agreement between the rankings with more usual measures, like Cohen's and Fleiss' kappa, or the correlation coefficients of Kendall and Spearman.
A particular care should however be given on the impact on these values of the unspecified agreements, produced by the presence of unordered pairs.

\paragraph{Acknowledgement} The authors is grateful to Itzel Morales-Ramirez for her help in executing the survey and her comments. This work is a result of the RISCOSS project, funded by the EC 7th Framework Programme FP7/2007-2013, agreement number 318249.

\bibliographystyle{cell}
\bibliography{citations}
\end{document}